\documentclass[a4paper,onecolumn,11pt,unpublished]{quantumarticle}
\pdfoutput=1
\usepackage[T1]{fontenc}
\usepackage[utf8]{inputenc}
\usepackage[english]{babel}
\usepackage[normalem]{ulem}
\usepackage{textcomp}
\usepackage{graphicx,float}
\usepackage{epstopdf}
\usepackage[dvipsnames,svgnames]{xcolor}
\usepackage{mathrsfs,mathtools,xfrac}
\usepackage{amsmath,amssymb,amsthm,amscd,bm}
\usepackage{multirow,comment}
\usepackage{array,booktabs,minibox}
\newcolumntype{P}[1]{>{\centering\arraybackslash}p{#1}} 
\usepackage[numbers]{natbib}
\usepackage[colorlinks,bookmarks=false,citecolor=blue,linkcolor=blue,urlcolor=blue,filecolor=blue]{hyperref}
\usepackage{wasysym}
\usepackage{makecell}

\newcommand{\anyon}[1]{\mathsf{#1}}
\newcommand{\rx}{\anyon{rx}}
\newcommand{\ry}{\anyon{ry}}
\newcommand{\rz}{\anyon{rz}}
\newcommand{\gx}{\anyon{gx}}
\newcommand{\gy}{\anyon{gy}}
\newcommand{\gz}{\anyon{gz}}
\newcommand{\bx}{\anyon{bx}}
\newcommand{\by}{\anyon{by}}
\newcommand{\bz}{\anyon{bz}}

\newcolumntype{C}{>{$}c<{$}}
\AtBeginDocument{
\heavyrulewidth=.08em
\lightrulewidth=.05em
\cmidrulewidth=.03em
\belowrulesep=.65ex
\belowbottomsep=0pt
\aboverulesep=.4ex
\abovetopsep=0pt
\cmidrulesep=\doublerulesep
\cmidrulekern=.5em
\defaultaddspace=.5em
}

\newcommand{\eref}[1]{Eq.~\eqref{#1}}
\newcommand{\sref}[1]{Sec.~\ref{#1}}
\newcommand{\fref}[1]{Fig.~\ref{#1}}
\newcommand{\tref}[1]{Table.~\ref{#1}}
\newcommand{\aref}[1]{Appendix.~\ref{#1}}

\begin{document}

\title{Spacetime Layout and Logical Compilation of Color Code}

\author{Qinjing Yu}
\affiliation{Hefei National Research Center for Physical Sciences at the Microscale and School of Physical Sciences, University of Science and Technology of China, Hefei 230026, China}
\affiliation{Hefei National Laboratory, University of Science and Technology of China, Hefei, 230088, China}
\affiliation{Shanghai Research Center for Quantum Science and CAS Center for Excellence in Quantum Information and Quantum Physics, University of Science and Technology of China, Shanghai 201315, China}
\email{qinjing.yu@mail.ustc.edu.cn}

\author{Ke Liu}
\affiliation{Hefei National Research Center for Physical Sciences at the Microscale and School of Physical Sciences, University of Science and Technology of China, Hefei 230026, China}
\affiliation{Shanghai Research Center for Quantum Science and CAS Center for Excellence in Quantum Information and Quantum Physics, University of Science and Technology of China, Shanghai 201315, China}
\email{ke.liu@ustc.edu.cn}

\begin{abstract}
Fault-tolerant quantum computing requires system-level coordination of logical primitives.
Here, we establish a logical compilation framework for the color code, grounded in its topological structure and supporting universal logical operations.
Based on its anyon-condensation and domain-wall structure, we introduce a spacetime block-diagram representation capturing logical patches and operations and derive the rules governing block assembly.
A correspondence with ZX diagrams further identifies the logical semantics of this representation and enables transformations that preserve the represented computation.
Moreover, we develop a code-derived compilation strategy that converts ZX representations of logical computations into valid color-code spacetime layouts. In this strategy, edge-decorated ZX diagrams tailor the logical representation to the color code under the block-assembly constraints, and fusion-region-aware routing exploits semantic equivalence during geometric embedding.
We automate the complete logical compilation process and demonstrate successful compilation across a broad range of algorithms.
Our work advances color-code architecture from individual primitives to the automated synthesis of logical computations, marking a significant step toward its full-stack quantum computing.
\end{abstract}

\maketitle

\section{Introduction}
Utility-scale fault-tolerant quantum computing (FTQC) is not merely an aggregation of logical qubits and logical gates.
To carry out a target computation, the abstract algorithm must be compiled into a carefully designed arrangement of logical primitives.
The resulting orchestration and resource requirements can differ substantially from a simple enumeration of logical patches and operations.
Hardware implementations proceed from the compiled structure rather than from the original logical-qubit and gate description.

This process is prominently reflected in the development of surface-code-based architectures~\cite{Dennis02,Fowler12,horsman_surface_2012}.
Alongside improvements in encoding and logical-gate strategies~\cite{Gidney25Yoked,Gidney24YBasis,Low26Denser}, the surface code has also seen sustained study and progressive refinement of its compilation schemes.
Two representative schemes are Clifford+$T$ compilation and Pauli-based compilation: the former preserves the Clifford+$T$ circuit structure while routing and scheduling logical primitives~\cite{Beverland22EDPC,LeBlond24CliffordT}, whereas the latter reformulates the computation as Pauli-product measurements, with Clifford effects absorbed into subsequent Pauli axes and final measurement bases~\cite{Litinski19Game,Litinski19Distillation}.
A further line of compilation is based on spacetime block diagrams and the ZX calculus~\cite{KissingerWetering2024Book}. In this scheme, logical computations are represented as networks organized across space and time, and compilation is recast as the construction and optimization of these networks.
Works along this direction include manually designed magic-state-factory layouts~\cite{Gidney19CCZ,Gidney19AutoCCZ}, exact synthesis of bounded logical subroutines~\cite{Tan24LaSsynth}, and heuristic compilation at the algorithm scale~\cite{Zhou26TopoLS}.

The two-dimensional (2D) color code is another distinctive topological code~\cite{Bombin06ColorCode} with growing experismental relevance~\cite{Lacroix25ColorCode}. 
Existing research has mainly explored its properties at the level of individual logical primitives, including transversal gates~\cite{kubica_universal_2015}, magic-state preparations~\cite{Gidney24Cultivation,Chamberland20Magic,Lee25Distillation}, lattice-surgery operations ~\cite{Landahl14Surgery,Thomsen24Parallel}, and decoding algorithms~\cite{Gidney23Circuits,Chubb21Tensor,Olle26Predecoder}.
Its logical compilation, nevertheless, remains largely unexplored; yet such capability is essential for the maturation of color-code-based FTQC.

To address this gap, we introduce a systematic framework for the logical compilation of the 2D color code. We introduce a spacetime block diagram as the intermediate representation of the compilation process, in which logical patches and logical operations are encoded as block graphs composed of prisms, pipes, and ports. 
The constructions of these elementary blocks manifest the richer anyon-condensation and domain-wall behavior of the color code relative to the surface code.
We then derive the assembly rules of these elementary blocks, including their connectivity rules and domain-wall matching rules. Transversal single-qubit Clifford gates, multi-patch logical Pauli measurements, and $\lvert T\rangle$-gate teleportation are all compactly encoded, realizing a universal set of logical operations. 
Moreover, we establish a correspondence between block diagrams and ZX diagrams, turning the geometric description into a semantic representation. The resulting ZX diagram encodes color-code-specific spider structures and fusion constraints and provides a semantics-preserving language for optimization. Thus, color-code computations become representable and transformable.

Next, we develop a code-derived compilation strategy and automate the procedure.
While the ZX formulation preserves semantics, it does not by itself produce a valid spacetime layout that conforms to the richer structural requirements of the color code. We introduce an edge-decorated ZX diagram and a fusion-region-aware routing scheme. The edge-decorated diagram simplifies the raw ZX diagram into a form tailored to the color code without violating the assembly rules, while the routing scheme exploits fusion equivalence to construct a valid layout with reduced spacetime volume. We implement this strategy as an automated procedure and demonstrate, across a wide array of algorithms, that it uniformly reduces logical spacetime volume relative to surface-code compilation.

Our framework moves color-code architecture from the construction of individual logical primitives to the automated synthesis of logical computations. 
It enables the code’s native logical capabilities to be coordinated across an entire algorithm; otherwise, the algorithm and QEC layers remain disconnected.
The framework preserves logical design freedom and accommodates the color code’s richer topological structure while maintaining semantic correctness throughout compilation.
The resulting specification is not tied to a particular circuit realization and is compatible with different physical implementations.

While completing this manuscript, we became aware of concurrent and independently developed work by Herzog et al.~\cite{Herzog26Pipe}, which introduces pipe diagrams for the triangular 6.6.6 color code.
The two representations share a common logical core---distance-independent prisms and pipes whose junctions reproduce ZX/Pauli-flow semantics---but use distinct block conventions and admissibility rules, and are therefore semantically compatible on their shared Clifford constructions rather than identical layout languages.
Herzog et al. use macroscopic compilation for ZX-to-pipe construction, illustrated through selected Clifford examples, and microscopic compilation for translating specified pipe diagrams into distance-dependent stabilizers, correlation surfaces, and syndrome-extraction circuits.
Here, logical compilation instead denotes an automated, algorithm-scale transformation of universal Clifford+$T$ circuits into validated block graphs through code-derived assembly rules and fusion-region-aware embedding, while physical-circuit realization is left to a downstream layer.

The remainder of this article is organized as follows. \sref{sec:spacetime-layout} develops the spacetime layout of the color code: after a brief review of the 2D color code, we introduce the building blocks of a spacetime layout, derive their assembly rules, and establish the correspondence to ZX diagrams. \sref{sec:compiling-strategy} presents the compiling strategy, built on the edge-decorated ZX diagram and fusion-region-aware routing. \sref{sec:demonstration} assembles these components into an automated compilation pipeline and benchmarks it across nine algorithms. \sref{sec:sum} summarizes and discusses future directions.

\section{Spacetime layout}\label{sec:spacetime-layout}

\subsection{Basics of 2D color code}

\begin{figure}[t]
  \centering
  \includegraphics[width=0.69\columnwidth]{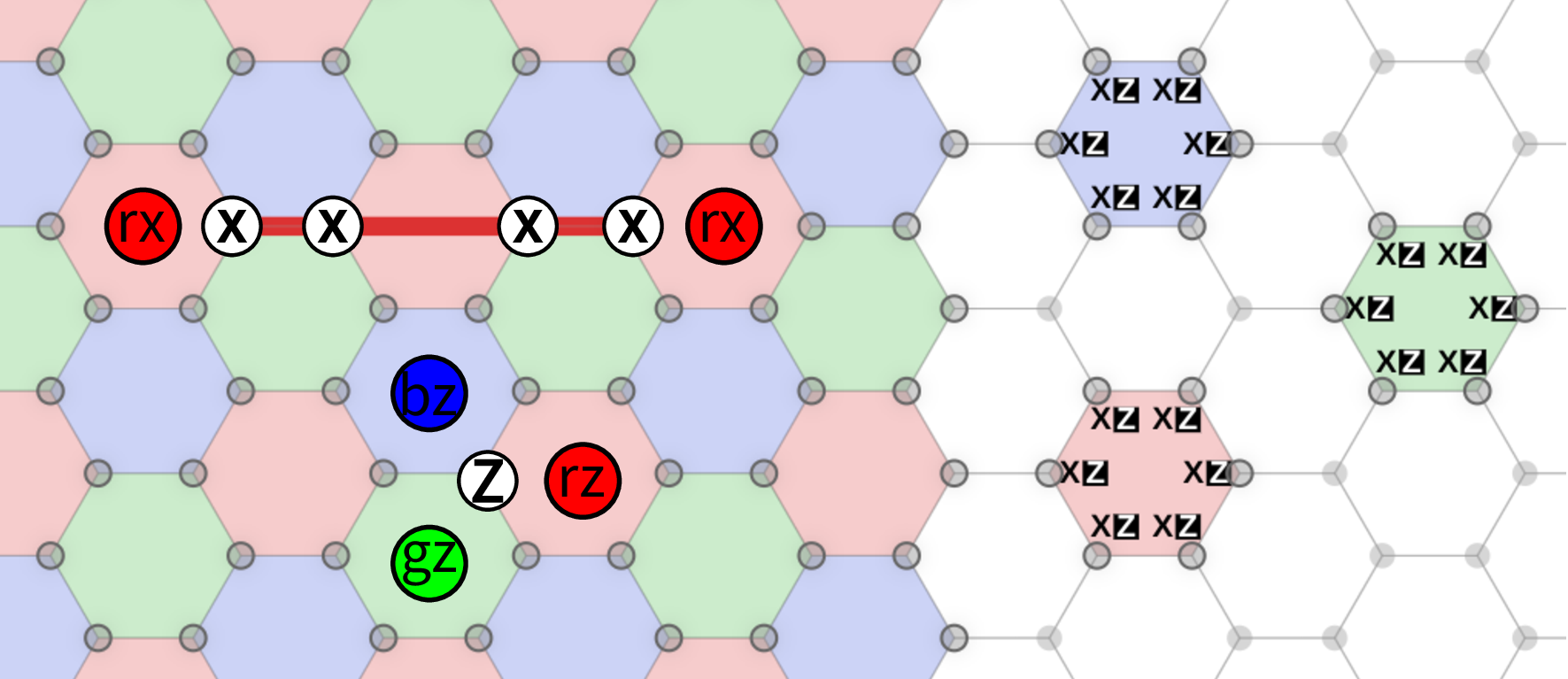}\hfill
  \includegraphics[width=0.29\columnwidth]{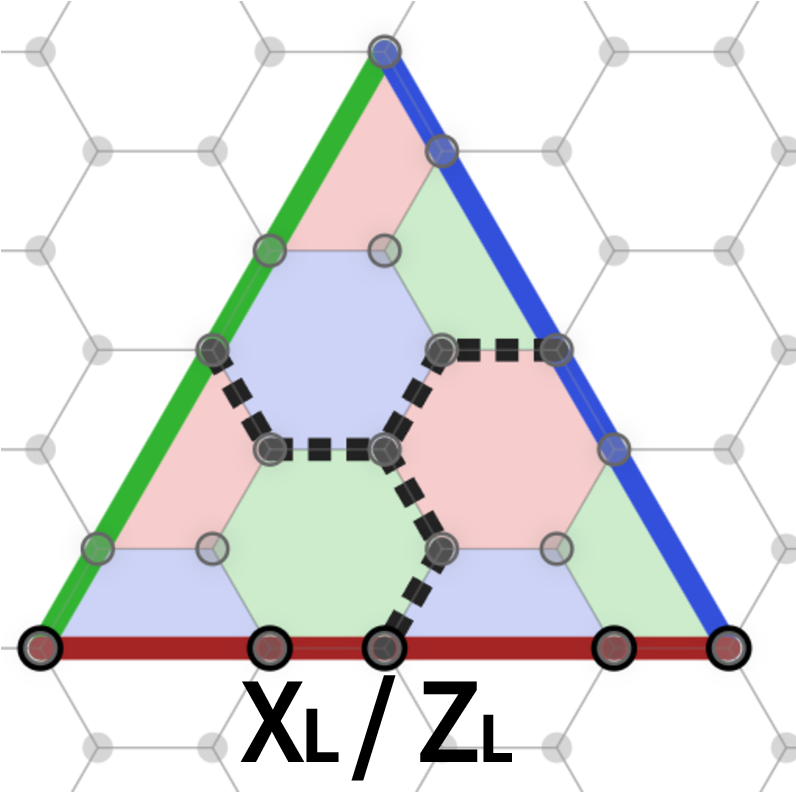}
  \caption{Left: The anyons in the color code. A string of physical $X$ operators supported on a path connecting two red faces creates a pair of $\rx$ anyons at those faces. A single physical $Z$ operator creates a triple of anyons $\rz$, $\gz$, and $\bz$ at the three faces adjacent to the qubit. Middle: The stabilizers of the color code. Each face hosts two stabilizers, one in the $X$ basis and one in the $Z$ basis. Right: A triangular patch of the 2D color code, with its three colour boundaries and logical operators. The logical Pauli operators can either be represented as star-shaped string operators in the bulk or as line operators along the boundaries.}
  \label{fig:triangular-patch-2d}
\end{figure}

The 2D color code is a topological stabilizer code defined on a lattice whose faces are three-colourable~\cite{Bombin06ColorCode}. In this work we focus on the hexagonal (6.6.6) lattice, which is hardware-friendly~\cite{Gidney23Circuits,Lacroix25ColorCode}. Each vertex of the lattice hosts a physical qubit, and each face $p$ hosts two stabilizer generators:
\begin{equation}
  s_X^p = X^{\otimes 6}, \quad s_Z^p = Z^{\otimes 6}
\end{equation}
acting in the $X$- and $Z$-basis on all qubits on the boundary of $p$, as shown in \fref{fig:triangular-patch-2d}. Their product yields a
$Y$-basis stabilizer, $s_Y^p=s_X^p s_Z^p$, so the code treats the three Pauli
bases on an equal footing.

A Pauli operator that violates some stabilizers creates anyons on the corresponding faces. The color code hosts 16 anyons: the vacuum, nine bosons, and six fermions~\cite{kesselringBoundariesTwistDefects2018,Kesselring24Condensation}. Each boson carries one of three colour labels, $\mathsf{r},\mathsf{g},\mathsf{b}$, the colour of the violated face, and one of three Pauli labels, $\mathsf{x},\mathsf{y},\mathsf{z}$, the basis of the Pauli string that creates and moves it. For example, a string of physical $X$ operators supported on a path connecting two red faces creates a pair of $\rx$ anyons at those faces, as shown in \fref{fig:triangular-patch-2d}. The nine bosons can be represented in the \emph{boson table}, in which bosons of one row share their Pauli label and bosons of one column share their colour label:

\[
\begin{array}{c|c|c}
  \rx & \gx & \bx \\ \hline
  \ry & \gy & \by \\ \hline
  \rz & \gz & \bz
\end{array}
\]

So far we have described the anyons of an infinite bulk. A finite code patch, however, must terminate at boundaries, and one may also join regions with different anyon behaviour along internal interfaces. Both are characterized by how they act on the anyons that reach them. Condensing a subgroup of bosons inside a subregion creates a \emph{domain wall} along its border: bosons of the condensed subgroup are absorbed by the domain wall, bosons that braid trivially with the whole subgroup pass through it (\emph{deconfined}), and all remaining anyons are \emph{confined}, which cannot cross or enter the domain wall. The action of a wall on the anyons arriving from one side is therefore summarized by marking each entry of the boson table as condensed, deconfined, or confined. A general domain wall acts differently on its two sides and is specified by one marked table per side, together with an identification of the deconfined bosons across the domain wall; the domain walls appearing in this work act identically on both sides, so we describe each by a single table.

Three types of domain walls occur in the color code. An \emph{opaque} domain wall condenses a maximal (Lagrangian) subgroup, i.e., a full row or a full column of the boson table, and confines everything else; the six such domain walls serve as the boundaries of the code to the vacuum, and we call them colour boundaries (columns) and Pauli boundaries (rows). A \emph{semi-transparent} domain wall condenses a single boson on one side, leaving the four bosons that braid trivially with it deconfined and the remaining four confined. A \emph{transparent} domain wall condenses nothing: every anyon crosses it, possibly with a permuted label, and such domain walls correspond one-to-one to the symmetries of the anyon model, which act on the boson table as row and column permutations and reflections about its diagonals.

Equipped with domain walls, we can terminate the lattice and encode logical information. In this work, we focus on logical qubits encoded in triangular regions of the lattice, whose three sides are red, green, and blue boundaries, as shown in \fref{fig:triangular-patch-2d}. A logical operator is a string operator along a topologically non-trivial path: it creates an anyon at one boundary and condenses it on other distinct boundaries. On the triangular patch a logical operator is generally star-shaped: for a Pauli label $p\in\{\mathsf{x},\mathsf{y},\mathsf{z}\}$, three strings transporting $\mathsf{rp}$, $\mathsf{gp}$, and $\mathsf{bp}$ meet at a fusion point in the bulk, and each leg terminates on the colour boundary that condenses it. Moving the fusion point onto one boundary contracts the leg of that colour to a point, and the star degenerates into a line connecting the remaining two boundaries. Its length $d$ is the code distance, which grows with the size of the patch. Because all three Pauli labels admit such representatives on the same patch, logical $X$, $Y$, and $Z$ are geometrically equivalent, a symmetry that the spacetime layout of the next sections inherits.

\subsection{Building blocks of spacetime layout}\label{sec:building-blocks}

\begin{figure*}[t]
  \centering
  \includegraphics[width=0.95\textwidth]{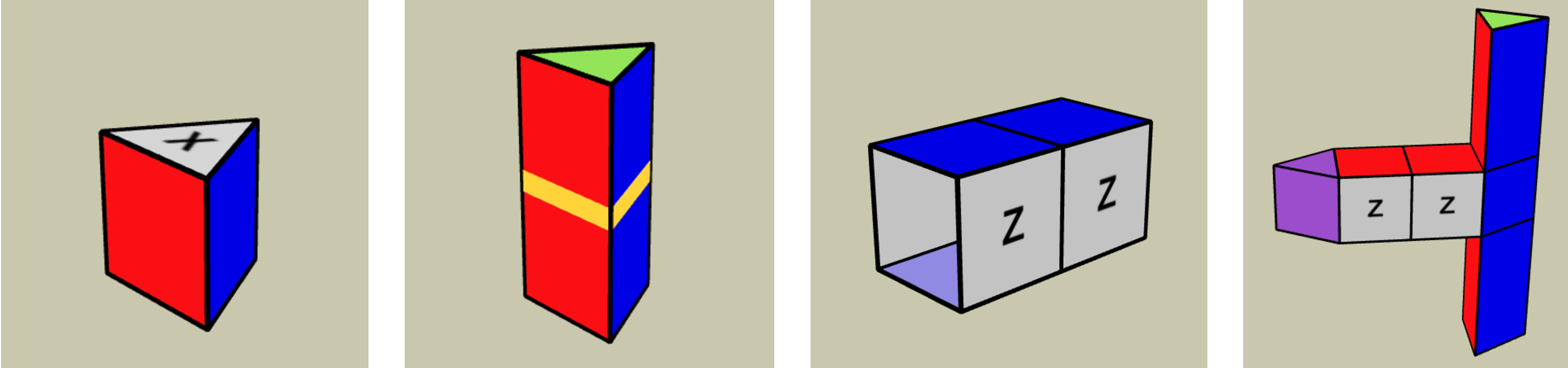}
  \caption{The building blocks of the spacetime layout of the 2D color code. Left: a prism, with its top and bottom faces as Pauli X boundaries and its lateral faces as colour boundaries. Middle left: a time pipe, with a transparent domain wall carrying a sequence of single-qubit Clifford gates. Middle right: a space pipe, with its top and bottom faces as blue colour boundaries and its lateral faces as Pauli Z boundaries. Right: a magic state injection port (purple), which is attached to a junction. A magic state is injected asynchronously at the port.}\label{fig:building-blocks}
\end{figure*}

So far we have described the color code on a static two-dimensional lattice. Fault tolerance, however, requires the stabilizers to be measured repeatedly, since the measurements themselves are noisy~\cite{Gidney23Circuits,Lacroix25ColorCode}. A patch under repeated measurement traces out a volume in $(2+1)$-dimensional spacetime, with its boundaries sweeping out the lateral faces. Initializing or reading out the patch terminates this volume in time, forming domain walls oriented in a temporal plane~\cite{Kesselring24Condensation}. A fault-tolerant computation is therefore an arrangement of volumes bounded by domain walls, which we call a \emph{spacetime layout}. This section defines its building blocks, as shown in \fref{fig:building-blocks}.

\textbf{Prism.} A prism is the spacetime trace of a triangular patch of distance $d$ maintained for $d$ rounds of syndrome measurement. This duration serves as the unit of time in the layout, and a prism occupies one unit of spacetime volume. Its top and bottom faces are Pauli boundaries, condensing one of the three Pauli families ($\mathsf{x}$, $\mathsf{y}$, or $\mathsf{z}$); a bottom (top) face of Pauli type $P$ realizes the initialization (readout) of the patch in the logical $P$ basis. Its three lateral faces are the traces of the patch's three colour boundaries, each condensing one colour family. The colours of the lateral faces of a standalone prism are arbitrary, but once the prism is incorporated into the layout, the colours of the connecting faces must match, see \sref{sec:assembly-rules}. 

\textbf{Time pipe.} A time pipe connects two vertically adjacent prisms. By convention, its bottom end is the input and its top end the output. Where a time pipe connects a prism, the shared horizontal faces are removed so that the patch simply persists in time instead of being read out and re-initialized. Like the prism, its lateral faces are colour boundaries. Although drawn with a finite extent, a time pipe adds no spacetime volume.

A time pipe can carry a transparent domain wall, which permutes the labels of crossing anyons and hence the logical operators, acting as a logical Clifford gate; preserving the lateral colour boundaries restricts the permutations to Pauli labels, i.e., up to phases, the single-qubit Clifford group generated by $H$ and $S$~\cite{Kesselring24Condensation}. We therefore label a transparent domain wall by an ordered sequence of single-qubit Clifford gates $(g_1,\ldots,g_n)$ with $g_i\in\{H,S\}$, applied in time order (earliest gate first, bottom to top). Because every single-qubit Clifford gate is transversal for the 2D color code~\cite{Bombin06ColorCode,kubica_universal_2015,Lacroix25ColorCode}, such a domain wall is realized within the syndrome-extraction rounds it interrupts: single-qubit Clifford gates are free in this representation.

\textbf{Space pipe.} A space pipe is a horizontal connection between two side-adjacent prisms. Inside the pipe a single bosonic anyon is condensed, so the pipe as a whole is a semi-transparent domain wall, and its interior is equivalent to the surface code: the four deconfined anyons crossing the pipe are transformed into the electric and magnetic anyons $e$ and $m$~\cite{Kesselring24Condensation,kubicaUnfoldingColorCode2015}.

A space pipe naturally divides into two halves, one adjacent to each prism, which we call \textbf{cubes}. The top and bottom faces of the two cubes can be chosen as either colour boundaries or Pauli boundaries of the same type. Both choices are equivalent for logical compilation, as detailed in \aref{app:junction-propagation}. Without loss of generality, we choose colour boundaries in this work, as shown in \fref{fig:building-blocks}. Each cube's two lateral faces are Pauli boundaries. Being boundaries of the surface-code interior, the top and bottom faces condense one of $e$ and $m$, and the lateral faces condense the other. Logical operators supported on the pipe are therefore anyon strings connecting opposing faces.

In general the two cubes may carry different lateral Pauli boundaries; the interface between them is then a transparent domain wall, and the pipe realizes a Clifford gate in the spatial direction once one endpoint is designated as input and the other as output. We deliberately exclude this possibility. Any Clifford gate placed in the spatial direction can be moved onto a time pipe as a transparent domain wall without adding spacetime overhead, whereas designating input and output endpoints in the spatial direction complicates compilation. We therefore restrict every space pipe to two identical cubes, whose lateral Pauli boundaries are of the same type.

\textbf{Port.} A port is a prism-free endpoint of a time pipe or a space pipe. Temporal ports on the bottom and top encode the logical inputs and outputs of the computation. Consistent with the restriction above, spatial ports are forbidden, with a single exception: a magic state injection port marks the open end of a designated space pipe that connects to an external logical patch, on which the magic state $(\lvert 0\rangle + e^{i\theta}\lvert 1\rangle)/\sqrt{2}$, for an arbitrary rotation angle $\theta$ about the $Z$ axis, is injected asynchronously by other components of the fault-tolerant system \cite{Zhou26TopoLS}. Magic state injection ports are the only non-Clifford ingredient of a layout, and everything else is Clifford. In this work we focus on Clifford+$T$ circuits, where every injection port injects the $\lvert T\rangle$ state with $\theta=\pi/4$, which is produced by components such as magic-state factories~\cite{Gidney19CCZ,Gidney19AutoCCZ}.

\subsection{Assembly rules}\label{sec:assembly-rules}

The building blocks cannot be attached arbitrarily: a valid spacetime layout requires well-formed connectivity and consistent domain walls at every attachment. In this section we enumerate the assembly rules that govern the construction of a spacetime layout based on the color code.

\textbf{Connectivity rules.} (i) \emph{No degree-one prism.} A prism attached to exactly one pipe is redundant and can always be pruned. (ii) \emph{No fanout port.} A port is the open end of exactly one pipe. (iii) \emph{No spatial port.} We forbid spatial ports except for magic state injection ports, as stated in \sref{sec:building-blocks}. An open end of a space pipe can always be bent into the temporal direction by appending a prism and a time pipe.

\textbf{Pauli matching.} All space pipes attached to the same prism must carry lateral Pauli boundaries of the same type. The lateral walls of two pipes on adjacent faces meet at the shared vertical corner of the prism, where the single-basis stabilizers realizing the two Pauli boundaries act on overlapping qubits; for two different Pauli types they fail to commute.

\textbf{Pauli overlap.} The lateral Pauli boundaries of a space pipe must differ in type from any exposed top or bottom face of the prism it attaches to, i.e., a face not replaced by a time pipe. Boundaries condensing the same boson family terminate the same anyon strings, and sharing one Pauli type would let a string take a shortcut from the lateral wall of the pipe, which breaks the fault tolerance.

\textbf{Colour matching.} The top and bottom faces of a space pipe must carry the colour of the prism faces at which it attaches, and the lateral colour boundaries of a time pipe must match those of the prisms it joins. A mismatched colour opens the same kind of shortcut.  Colour matching at both ends of a space pipe requires the two endpoint prisms to present the same colour at their shared interface, which we guarantee globally by construction.

\begin{figure*}[t]
  \centering
  \includegraphics[width=0.48\textwidth]{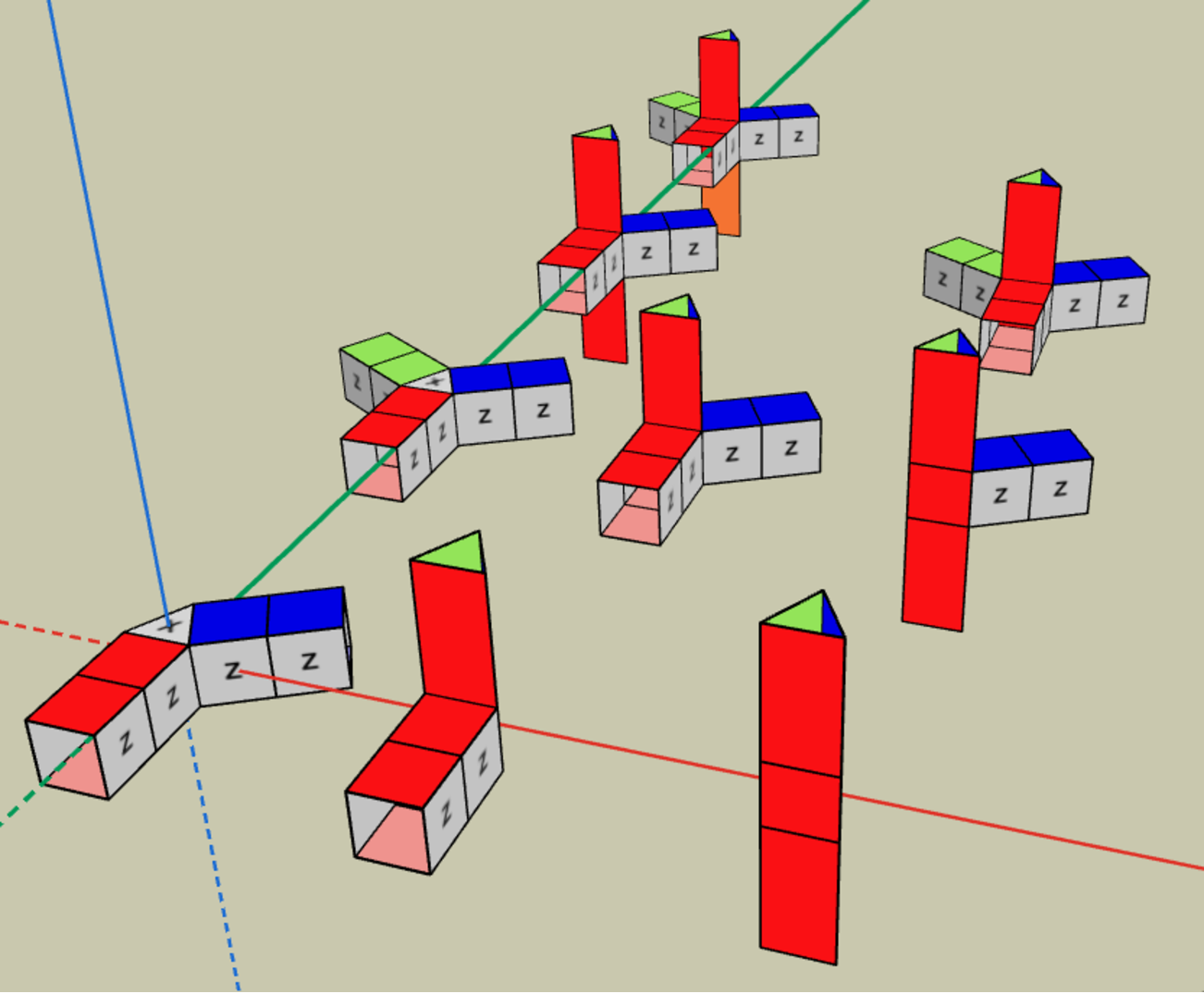}\hfill
  \includegraphics[width=0.48\textwidth]{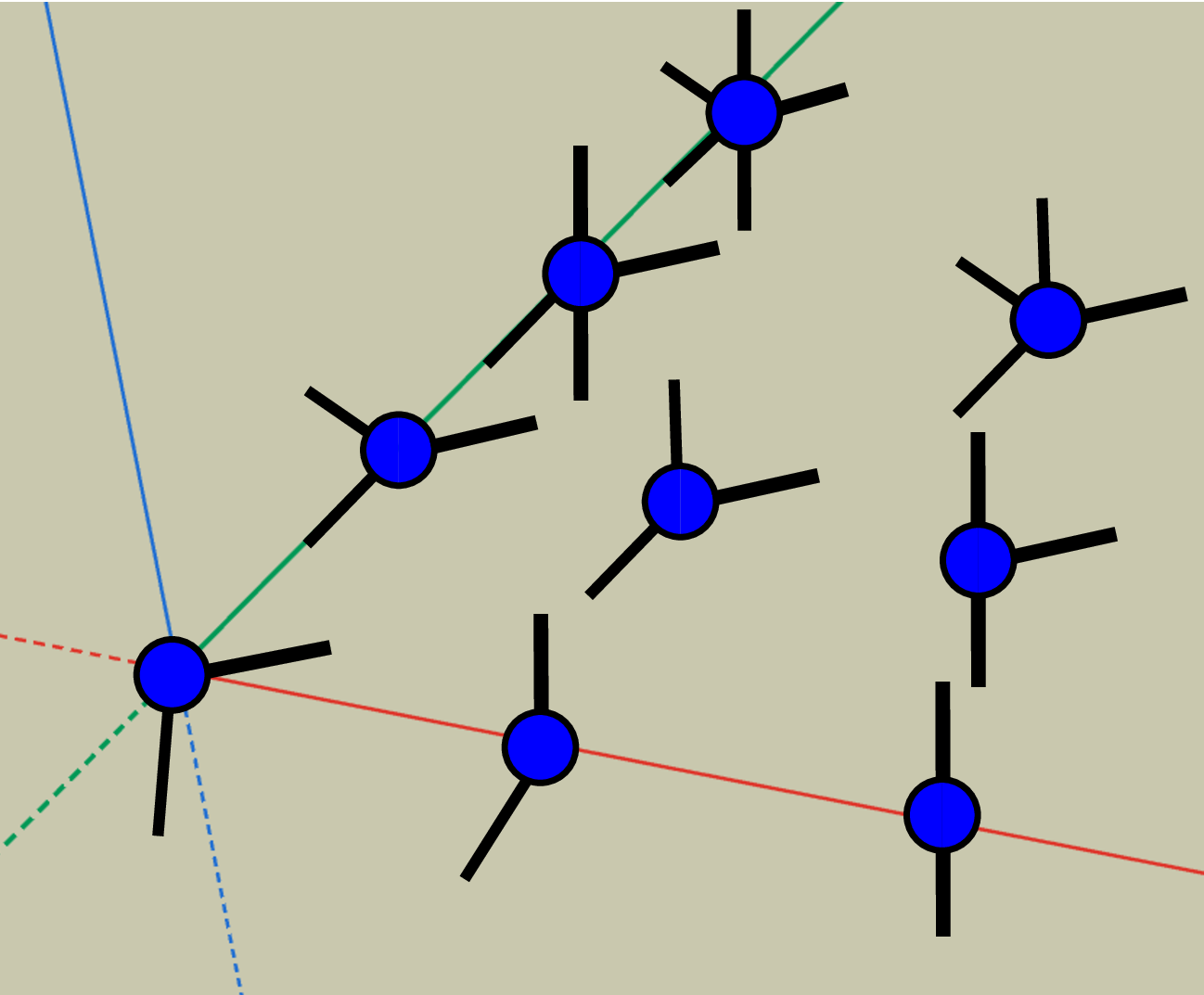}
  \caption{Left: All possible z-basis junctions of degree 2, 3, 4, and 5 in the spacetime layout of the 2D color code. Right: The corresponding ZX diagram of each junction.} \label{fig:junctions-z-basis}
\end{figure*}

Under these rules, a prism can host at most two time pipes and three space pipes simultaneously, and forms a \textbf{junction} of degree up to five. In \fref{fig:junctions-z-basis} we enumerate all possible junctions of degree 2, 3, 4, and 5 with lateral Pauli Z boundaries, and junctions with lateral Pauli X boundaries follow the same structures. By contrast, a surface-code cube supports at most degree four: pipes along all three axes would force a boundary-type conflict at a corner, so at most two of its axes carry pipes~\cite{Zhou26TopoLS,Tan24LaSsynth}. This higher junction degree allows a layout to realize the same connectivity with fewer junctions, which is a key factor in reducing the spacetime volume of a logical circuit based on the color code.

\subsection{Correspondence to ZX diagram}\label{sec:correspondence-zx}

A structurally valid spacetime layout implements a logical computation. For the surface code, such a computation is naturally expressed in the ZX calculus: the merges and splits of lattice surgery realize exactly the spider tensors, so a protocol can be read as a ZX diagram~\cite{Zhou26TopoLS,beaudrapZXCalculusLanguage2020}. In this section we establish the corresponding description for the color code: each building block introduced above has an image in a ZX diagram, so that every layout assembled from these blocks translates into a ZX diagram representing its logical computation.

A ZX diagram is a graph whose vertices are spiders of two species, Z spiders and X spiders, connected by wires, with open wire ends being the inputs and outputs of the diagram. Each spider carries a phase $\alpha$ and an arbitrary number of legs. Specifically, an S gate can be represented as a $Z(\pi/2)$ spider with one input and one output, a T gate as a $Z(\pi/4)$ spider with one input and one output. The Hadamard gate is a distinguished degree-2 box. Crucially, a ZX diagram specifies only the connectivity between spiders: wires may be bent, stretched, and rearranged freely without changing the linear map the diagram represents~\cite{KissingerWetering2024Book,townsend-teagueFloquetifyingColourCode2023,Zhou26TopoLS}.

For the correspondence below we only need how Pauli operators propagate through a phase-0 spider. A phase-0 X spider with $n$ legs is characterized by two families of stabilizers of its tensor: $X_iX_j$ for any pair of legs $i,j$, and $Z^{\otimes n}$ over all legs. Consequently, Pauli $X$ operators enter and leave the spider on any even number of legs (the \emph{parity rule}), while a Pauli $Z$ operator must occupy either all legs or none (the \emph{all-or-nothing rule}). The rules for a Z spider are obtained by exchanging $X$ and $Z$. A consistent assignment of Pauli operators to the wires of a diagram is called a \emph{Pauli web} \cite{bombinUnifyingFlavorsFault2024, PRXQuantum_xyz_pauliflow}.

In this correspondence, pipes become the wires of the ZX diagram and prisms become its spiders. Which spider a prism realizes is a property of its junction, determined by how logical Pauli operators propagate through it. For a junction whose space pipes all carry lateral Pauli X boundaries, deforming the star-shaped string operators through its domain walls shows that a logical X operator entering from one pipe leaves on exactly one other pipe. Products of such deformations realize every configuration in which an even number of pipes carries a logical X. By contrast, a logical Z operator entering from one pipe leaves on all the other pipes simultaneously. These are precisely the parity rule and the all-or-nothing rule of a phase-0 X spider, so the junction corresponds to an X spider of the same degree. A junction with lateral Pauli Z boundaries likewise corresponds to a Z spider, as enumerated in \fref{fig:junctions-z-basis}. Because a colour boundary treats the three Pauli labels symmetrically while a ZX diagram does not, the anyon strings in the space pipes must be identified with a definite Pauli label by convention. The derivation and this identification are detailed in \aref{app:junction-propagation}.

So far the time pipes were bare. A time pipe carrying a transparent domain wall $(g_1,\ldots,g_n)$ trivially corresponds, on the ZX side, to the same ordered sequence placed along the temporal wire: each $H$ becomes a Hadamard box and each $S$ becomes a $Z(\pi/2)$ spider, in time order. Since the domain wall occupies a wire rather than a vertex, prisms remain phase-0 spiders and the correspondence established above is unaffected.

Finally, the ports. A temporal port is an open wire end of the ZX diagram: lower ports are inputs and upper ports are outputs. A magic state injection port corresponds to a degree-1 $Z(\theta)$ spider attached to the wire of its space pipe, a $Z(\pi/4)$ spider in this work. Every element of a spacetime layout now has a ZX image, and the correspondence extends to the entire layout by composition. The resulting ZX diagram is a semantic representation of the logical computation implemented by the layout.

A ZX diagram can be transformed by the rewrite rules of the ZX calculus with the represented linear map preserved. The correspondence lifts this freedom to spacetime layouts: any valid layout realizing the transformed diagram implements the same computation. Compilation can therefore operate on the diagram rather than on the geometry, applying semantics-preserving transformations before embedding the result back into spacetime, as the next section develops. The diagram also supplies the correctness criterion for the whole pipeline: a compiled layout is correct precisely when its ZX diagram is equivalent to the ZX diagram of the logical circuit to be implemented.

\section{Compiling strategy}\label{sec:compiling-strategy}

The preceding section established the spacetime layout of the color code and its correspondence to ZX diagrams. We now turn to logical compilation: given an abstract logical circuit, the task is to construct a valid spacetime layout that implements it, with the spacetime volume kept as small as possible.

\subsection{Edge-decorated ZX diagram for the color code}\label{sec:edge-decorated-zx}

Compilation starts from the ZX diagram directly translated from the input circuit. We rewrite it into a form matched to the building blocks of the layout, the \textbf{edge-decorated ZX diagram}: a diagram whose vertices are $Z(0)$ spiders, $X(0)$ spiders, and boundaries, together with a decoration assigning some edges an ordered sequence of $H$ and $Z(\pi/2)$ as S gates. By the correspondence of \sref{sec:correspondence-zx}, every element of this form has a direct layout image. The assembly rules bound the connectivity of this form: a prism hosts at most five pipes, so every vertex must have degree at most 5. Three rewriting steps bring the raw ZX diagram into this form.

\textbf{Spider fusion.} The raw ZX diagram scatters its connectivity over many low-degree spiders, each of which would consume a prism. Fusing spiders concentrates it onto fewer spiders of higher degree, exploiting the degree-5 junctions of the color code. The procedure is a single greedy sweep of fusion together with the Hadamard-pair cancellations, Hopf rule, and identity removals that expose further fusions~\cite{KissingerWetering2024Book}.

This simplification step acts on whole \textbf{fusion regions} rather than only adjacent same-kind pairs. A fusion region is the maximal set of spiders that can fuse into one without touching the rest of the diagram, and comes in two types. Two X spiders belong to one fusion region of type X exactly when the degree-two vertices between them compose to a rotation about the X axis, with canonical forms $(I)$, $(H, Z(\pi/2), H)$, $(H, Z(\pi/2), Z(\pi/2), H)$, and $(Z(\pi/2), H, Z(\pi/2))$ realizing $X(0)$, $X(\pi/2)$, $X(\pi)$, and $X(3\pi/2)$. Likewise, two Z spiders belong to one fusion region of type Z exactly when the vertices between them compose to a rotation about the Z axis, canonically $(I)$ or a single $Z(\pi/2)$, $Z(\pi)$, or $Z(3\pi/2)$ spider. The interior then collapses into the merged spider, contributing its rotation angle to the phase.

Each surviving spider will be eventually realized by a junction in the spacetime layout, which hosts at most five pipes. The degree of a merged spider is therefore bounded by five: a fusion is skipped when it would exceed this bound, and the bound tightens to four for a Z spider whose phase keeps a $\pi/4$ residue, since the next step will grow one more leg on such a spider.

\textbf{Phase normalization.} Spider fusion leaves the phases of the circuit concentrated on the surviving spiders, but the layout offers only two carriers of such phases: transparent domain walls in time pipes realize sequences of $H$ and $S$ gates, and $\lvert T\rangle$ ports supply the $\pi/4$ rotations. Phase normalization rewrites every phased spider into exactly these forms. A spider of phase $m(\pi/2)$ keeps its kind and drops to phase 0, shedding the phase onto one incident wire as a chain of degree-two vertices that composes to the rotation about its own axis. A Z spider emits $m$ consecutive $Z(\pi/2)$ vertices; an X spider emits the canonical form of the corresponding X rotation listed above. A Z spider of phase $m(\pi/2)+\pi/4$ sheds its Clifford part the same way and additionally grows a fresh leg ending in a degree-one $Z(\pi/4)$ spider. As a boundary case, a spider with a non-zero phase that is already of degree two and Clifford stays untouched without unfusing extra spiders.

\textbf{Edge decoration.} After phase normalization, all Clifford phases take the form of degree-two $H$ or $Z(\pi/2)$ vertices between phase-0 spiders or boundaries, and every $\pi/4$ phase sits on a degree-one $Z(\pi/4)$ spider. The edge decoration step extracts both from the diagram. Between each pair of phase-0 spiders or boundaries joined through such degree-two vertices, it reads the vertices in order as a sequence of $H$ and $S$ gates and normalizes the sequence to the canonical form of the single-qubit Clifford gate it composes to, which is one of the 24 elements of the single-qubit Clifford group (ignoring global phases). Then we replace the vertices by a single edge decorated with that sequence, connecting the pair directly. Meanwhile, each degree-one $Z(\pi/4)$ spider drops its phase and becomes a boundary vertex labelled as a $\lvert T\rangle$ state injection port. The resulting diagram is completely phase-free. The decoration is where the form is tailored to the color code: single-qubit Clifford gates are transversal, so a decorated edge embeds into one time pipe whose stacked transparent domain walls realize its sequence, at no additional spacetime volume.

The result is the edge-decorated ZX diagram of the input circuit: a completely phase-free skeleton in which every vertex fits a junction and every element has its spacetime layout image. What remains is geometry: assigning coordinates and routes so that these images assemble into a valid spacetime layout, the subject of the next section.

\subsection{Fusion-region-aware routing}\label{sec:fusion-routing}

Routing embeds the edge-decorated ZX diagram into the prism grid. Every spider is placed on a prism, and every edge is realized as a route, a path of prisms connected by pipes that starts and ends at the prisms or ports of its two endpoints. A decorated edge with a non-zero phase must route through at least one time pipe, in which its gate sequence is realized as stacked transparent domain walls; when the shortest path is purely spatial, the router takes the shortest detour through a time pipe instead.

The problem of routing has considerable freedom, and how this freedom is used is essential for reducing the spacetime overhead. We observe that the prisms along a route are essentially degree-two spiders unfused from the original $Z(0)$ and $X(0)$ spiders of the edge-decorated ZX diagram, working merely as idle connections. Each such prism extends the fusion region of the spider it is unfused from: in the spacetime layout, the fusion region of a placed spider is the maximal set of prisms connected to its placement prism through pipes by the criterion of \sref{sec:edge-decorated-zx}. Since every prism of a fusion region realizes the same role in the ZX diagram, a route for the decorated edge may start and end at any prism of the two regions rather than at the two placement prisms. Routes thereby shorten, and the extra prisms in earlier routes are reused by later ones.

\begin{figure}[t]
  \centering
  \includegraphics[width=0.62\columnwidth]{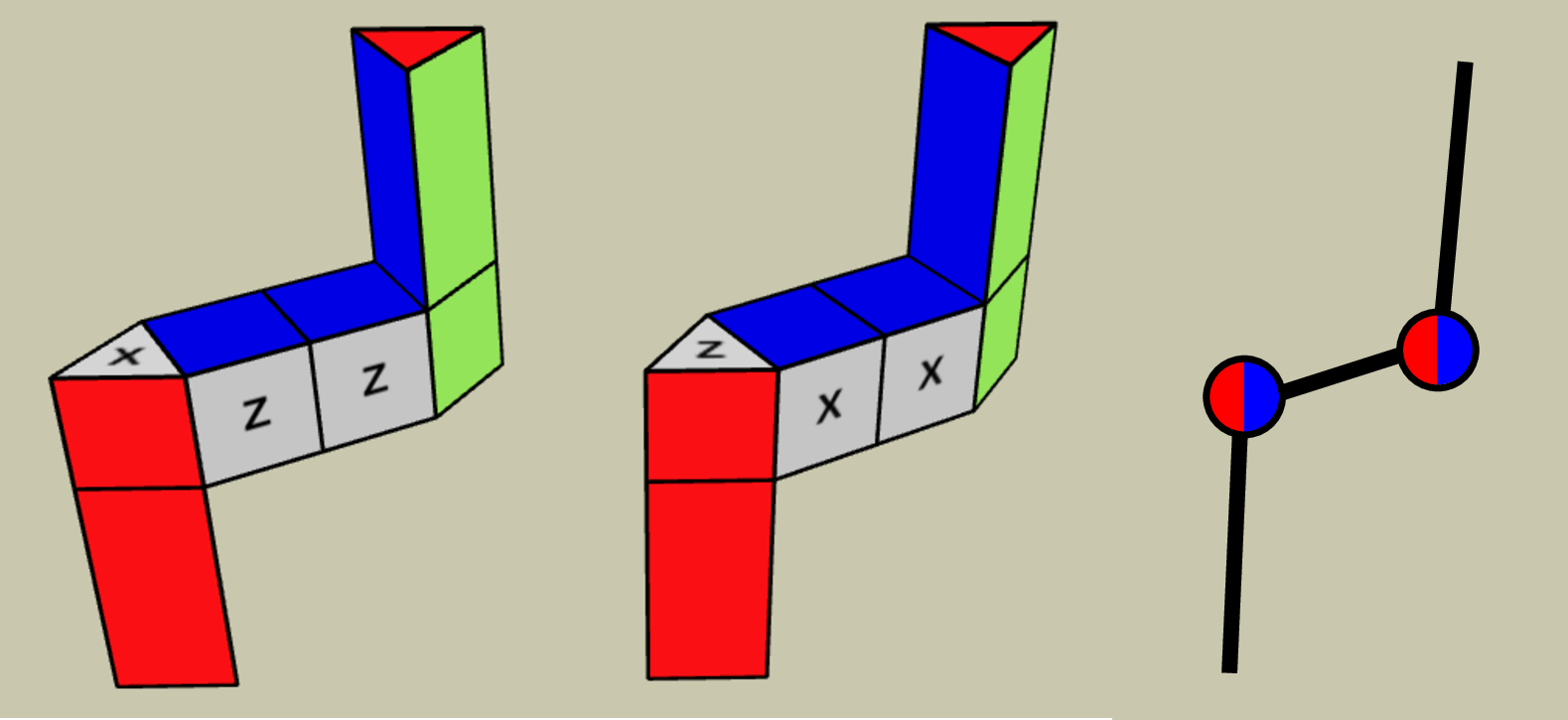}\hfill
  \includegraphics[width=0.30\columnwidth]{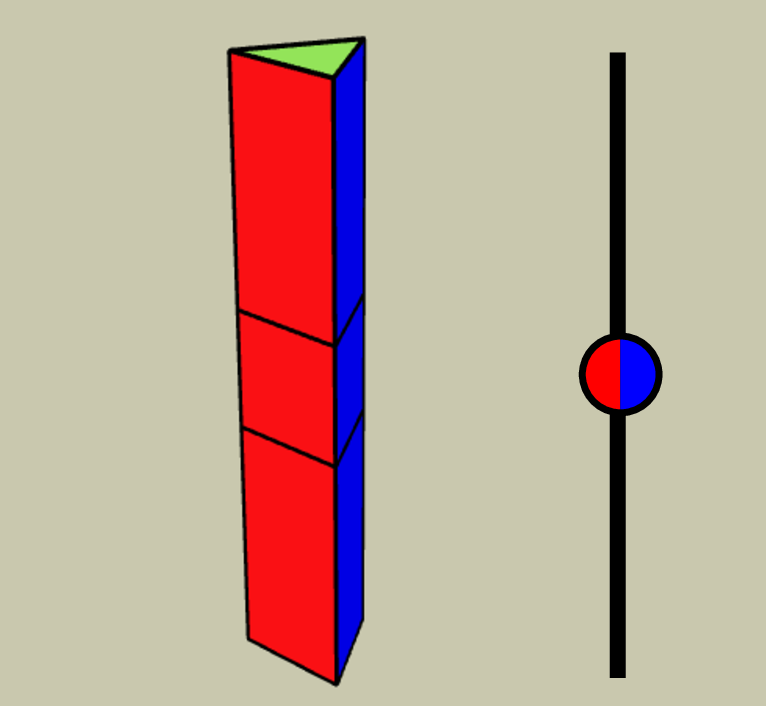}
  \caption{Left: Two prisms with deferred basis in the spatial direction, and the corresponding representation in ZX diagram. Right: One prism with deferred basis in the temporal direction, and the corresponding representation in ZX diagram.}
  \label{fig:deferred-spider}
\end{figure}

The basis of an interior prism through a route is, in principle, arbitrary, because degree-two $Z(0)$ and $X(0)$ spiders are equivalent in the ZX diagram by the identity rule. Therefore, the router keeps it deferred until another route claims the prism, which then adopts the kind of the claiming region, as shown in \fref{fig:deferred-spider}. By the Pauli matching rule in \sref{sec:assembly-rules}, space pipes share the lateral Pauli basis of the prisms they attach to, so a spatial stretch bounded by time pipes defers and resolves collectively. Time pipes impose no basis constraint, so in the temporal direction the deferral is per prism. A deferral that no route ever claims is pure wire, and its basis is fixed arbitrarily when the layout is finalized.

Attaching routes inside fusion regions can strand prisms: when the connections of a spider all arrive through other prisms of its region, its placement prism, or a prism of an earlier route, may end up attached to a single pipe, violating the no-degree-one-prism rule of \sref{sec:assembly-rules}. Such a leaf is a spider unfused from the region of its neighbour, so pruning it together with its pipe merely fuses it back, preserving the linear map. A transparent domain wall on the deleted pipe is re-anchored by composition onto another time pipe of the fusion region, and the leaf is retained if none exists. A logical spider on a pruned leaf moves to the surviving neighbour, and pruning iterates to a fixpoint, skipping ports and prisms still awaiting connections.

Fusion-region-aware routing fuses or unfuses spiders in the geometry, so a routed layout is equivalent to the input diagram by construction. The spacetime overhead reduction is quantified in \sref{sec:demonstration} against the baseline without fusion-region awareness, in which every route terminates at the two placement prisms.

\section{Demonstration of logical compilation}\label{sec:demonstration}

We now assemble the preceding components into an automated compilation pipeline that compiles a Clifford+$T$ circuit into a valid spacetime layout.

The input circuit is first cut in circuit time into blocks: each block is compiled independently, and the resulting layouts are stacked in time, with the output ports of one block meeting the input ports of the next on a shared interface plane. A cut through circuit time is crossed by exactly one worldline per qubit, so blocks always meet through one port per qubit.

Each block is translated into a ZX diagram and rewritten into its edge-decorated form by the three steps of \sref{sec:edge-decorated-zx}. A breadth-first traversal then slices the edge-decorated ZX diagram into layers, fixing the order in which spiders enter the search. A wire that crosses a layer without hosting a spider receives an idle spider, a phase-free degree-two spider that carries the wire through the layer and whose basis is deferred as in \sref{sec:fusion-routing}. The layering also determines where the circuit is cut: a spider whose connections continue into later layers keeps one port on the plane where its layer ends, and a plane hosts one port per qubit, so a block is closed before any layer of its diagram carries more such spiders than logical qubits.

The layers are embedded in sequence. We choose Monte Carlo tree search (MCTS) as the algorithm for layer embedding, for direct comparison with the existing work~\cite{Zhou26TopoLS}. The strategy of \sref{sec:compiling-strategy}, however, enters the pipeline only through the diagram rewriting and the routing, not through the search itself, and thus composes with any compilation algorithm that constructs a spacetime layout from a ZX diagram. A node of the search tree is a partially constructed layout; a move places one spider of the current layer on a prism near its already placed neighbours and routes the connecting edges by fusion-region-aware routing, realizing decorated edges on time pipes along the routes. The reward of a completed layer is the negative of its spacetime volume, steering the search toward compact embeddings. Once a layer is selected, the layout is pruned as described in \sref{sec:fusion-routing}, and the committed geometry becomes the fixed foundation of the next layer. 

\begin{figure}[t]
  \centering
  \includegraphics[width=0.48\columnwidth]{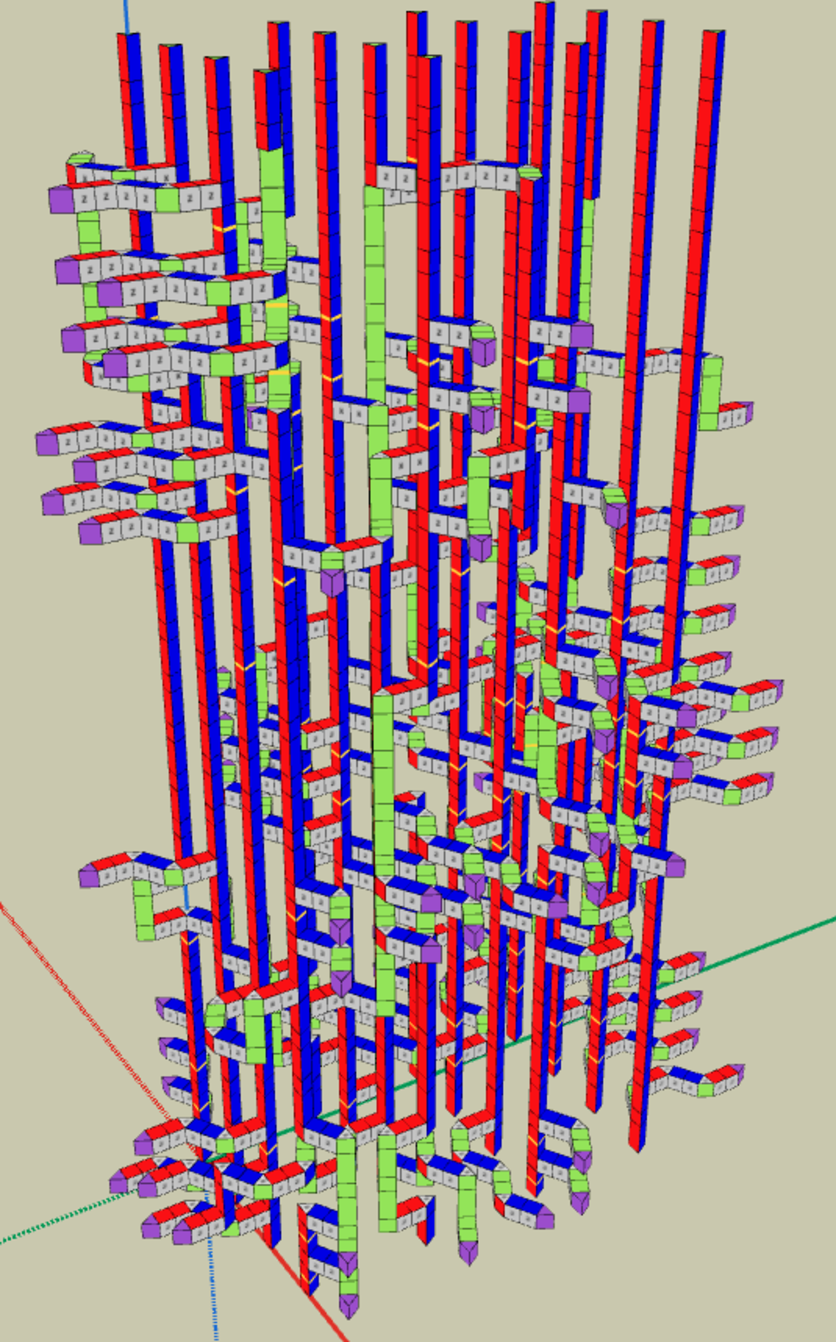}\hfill
  \includegraphics[width=0.48\columnwidth]{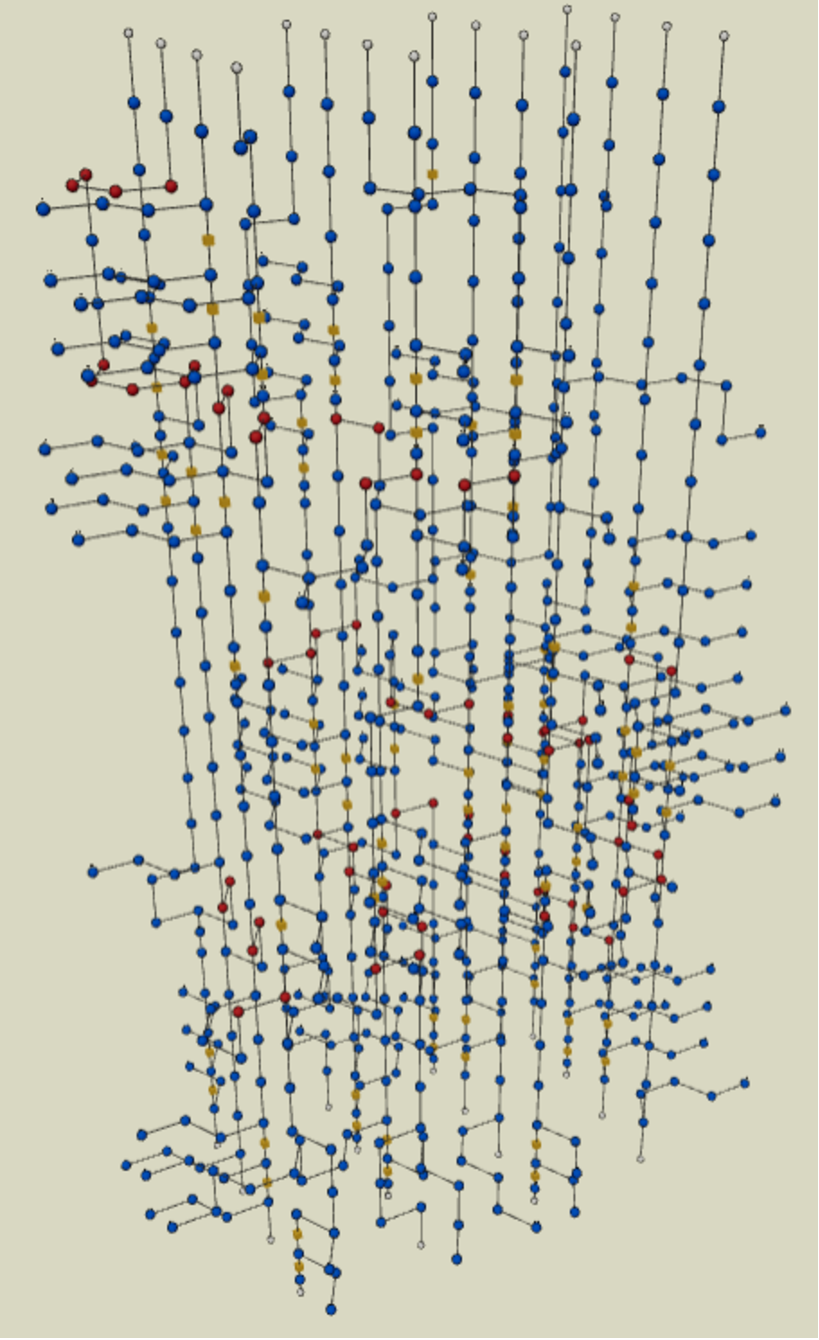}
  \caption{The result of our compilation pipeline for the 16-qubit VQE circuit. Left: the compiled spacetime layout. Right: the corresponding ZX diagram.}
  \label{fig:vqe16-layout-zx}
\end{figure}

A layer that fails to embed triggers a recovery cascade. The search first retries with a widened set of candidate moves and additional restarts; if the block still fails, it is bisected in circuit time and the halves are compiled independently. The bisection is always legal, since any cut through circuit time yields the same one-port-per-qubit interface. The recursion bottoms out at blocks that can be cut no further, which are handled by a deterministic fallback builder that always succeeds at the price of a larger volume and no fusion-region awareness. Finally, the block layouts are stitched on their shared interface planes, and the assembled layout is validated against the assembly rules of \sref{sec:assembly-rules}; the pipeline returns only layouts that pass this validation. \fref{fig:vqe16-layout-zx} shows an example of the compilation: the spacetime layout of the 16-qubit VQE circuit, together with its ZX diagram.

We compare the compilation results of TopoLS and our pipeline using two configurations. For our pipeline, the occupied-prism and bounding-box volumes are reported separately. For each circuit and configuration, we select the successful run with the smallest bounding-box volume among eight seeds, breaking ties by occupied-prism volume. For TopoLS, we use the smallest bounding-box spacetime volume reported for all configurations in Table~2 (Overall Comparison) of the TopoLS paper~\cite{Zhou26TopoLS}.

With fusion-region-aware routing, our pipeline attains a smaller bounding-box volume than TopoLS on all nine circuits, as shown in \tref{tab:benchmarking-comparison}. The comparison between the two configurations quantifies the contribution of fusion-region-aware routing: it reduces both volume metrics on every circuit relative to the baseline without it. The occupied-prism volume, which counts only the prisms actually used, lies well below the bounding-box volume, indicating additional headroom for compaction.
\begin{table}[t]
  \centering
  \caption{Comparison of compilation results. TopoLS reports bounding-box spacetime volume, while we report occupied-prism and bounding-box volumes in separate rows. ``Fusion'' refers to fusion-region-aware routing, while ``inplace'' refers to routing without fusion-region awareness.}
  \label{tab:benchmarking-comparison}
  \begin{tabular}{lrrr}
      \toprule
      Compilation pipeline / metric & BV & DJ & Grover \\
      \midrule
      TopoLS (bbox) & 486 & 891 & 23065 \\
      This work (inplace, prism) & 322 & 536 & 7079 \\
      This work (inplace, bbox) & 728 & 1458 & 19495 \\
      \textbf{This work (fusion, prism)} & \textbf{113} & \textbf{120} & \textbf{5247} \\
      \textbf{This work (fusion, bbox)} & \textbf{224} & \textbf{243} & \textbf{16660} \\
      \midrule
      & QFT & QPE & VQE \\
      \midrule
      TopoLS (bbox) & 36531 & 39447 & 4212 \\
      This work (inplace, prism) & 13128 & 13725 & 1229 \\
      This work (inplace, bbox) & 40824 & 42444 & 3726 \\
      \textbf{This work (fusion, prism)} & \textbf{8729} & \textbf{9288} & \textbf{682} \\
      \textbf{This work (fusion, bbox)} & \textbf{24705} & \textbf{26973} & \textbf{1701} \\
      \midrule
      & GHZ State & W State & QAOA \\
      \midrule
      TopoLS (bbox) & 243 & 8505 & 4374 \\
      This work (inplace, prism) & 87 & 1702 & 1961 \\
      This work (inplace, bbox) & 189 & 6075 & 5589 \\
      \textbf{This work (fusion, prism)} & \textbf{68} & \textbf{1405} & \textbf{1179} \\
      \textbf{This work (fusion, bbox)} & \textbf{126} & \textbf{5184} & \textbf{2916} \\
      \bottomrule
    \end{tabular}
\end{table}

\section{Summary and Outlook} \label{sec:sum}

We establish a systematic framework that enables automated logical compilation of the color code from its individual logical primitives. The framework incorporates the code’s topological structure into both the abstract description of logical operations and the construction of admissible spacetime layouts. It provides a tractable basis for system-wide optimization of color-code-based quantum computation.

The topological structure of the color code determines a spacetime block-graph representation and the rules by which its elements can be assembled. In this representation, prisms encode logical patches, pipes encode their spacetime connections, and ports specify inputs, outputs, and magic-state injections. The resulting block language supports universal computation through transversal Clifford gates, multi-patch Pauli measurements, and $T$-gate teleportation. Its assembly rules preserve domain-wall compatibility, enforcing colour and Pauli matching while allowing junctions of degree up to five. Moreover, by propagating star-shaped logical Pauli operators through the domain walls, we establish a ZX correspondence that assigns a spider to each junction.
This correspondence identifies semantically equivalent color-code block structures, enabling transformations that preserve the represented logical computation.

The compilation strategy further turns this semantic freedom into geometric freedom for constructing admissible spacetime layouts. We first transform the target algorithm into an edge-decorated ZX diagram that maps directly onto the block language. Fusion-region-aware routing then embeds entire fusion regions rather than fixed representative prisms, allowing later routes to attach to and reuse compatible geometry created by earlier routes. The automated procedure preserves the logical map through ZX rewriting, fusion, and unfusion, while validating the completed block graph against the assembly rules. We demonstrate the resulting pipeline on nine algorithmic benchmarks. Fusion-region-aware routing substantially reduces logical-layout volume relative to routing without it and consistently yields more compact layouts than the reported surface-code counterparts. These successful compilations confirm that the pipeline operates across a broad range of computational tasks.

Our work opens several directions for future research.
The block-diagram representation and fusion-region-aware embedding are independent of the particular search engine. The framework can therefore incorporate alternative exact, heuristic, or hybrid search methods to explore tradeoffs between search scalability and layout optimality. This common interface also enables systematic comparisons of search strategies across algorithmic tasks with different structures.
Moreover, the compiled logical specification can accommodate different color-code boundary constructions~\cite{Kesselring24Condensation,Lee25Distillation} and syndrome-extraction realizations~\cite{Gidney23Circuits,Chamberland20Flags}. This integration would enable logical compilation and downstream physical-circuit compilation to be designed and optimized jointly, advancing toward full-stack color-code FTQC.
More broadly, our framework may be extended to QLDPC codes such as bivariate-bicycle codes~\cite{Bravyi24Memory}, which admit topological descriptions~\cite{Chen25Anyon,Liang25TwistedTori} and open-boundary realizations~\cite{Steffan25TileCode}. Logical compilation of such codes is expected to be more challenging, but their topological structure may provide a principled route to developing code-derived compilation strategies.

\acknowledgments{}
This work is supported by the Strategic Priority Research Program of Chinese Academy
of Sciences (Grant No. XDB1680000).
The code and data will be made publicly available.
During the preparation of this manuscript, we became aware of concurrent and independent work on color-code compilation~\cite{Herzog26Pipe}.
While the two works share similar foundations in block-diagram and ZX representations, they address different compilation scopes. Herzog et al. demonstrate compact color-code layouts and physical-circuit realizations for selected Clifford operations, whereas we establish an automated, algorithm-scale logical compilation framework for universal Clifford+$T$ computations.

\appendix

\section{Propagation of logical operators through a junction}\label{app:junction-propagation}

\begin{figure*}[t]
  \centering
  \includegraphics[width=0.9\textwidth]{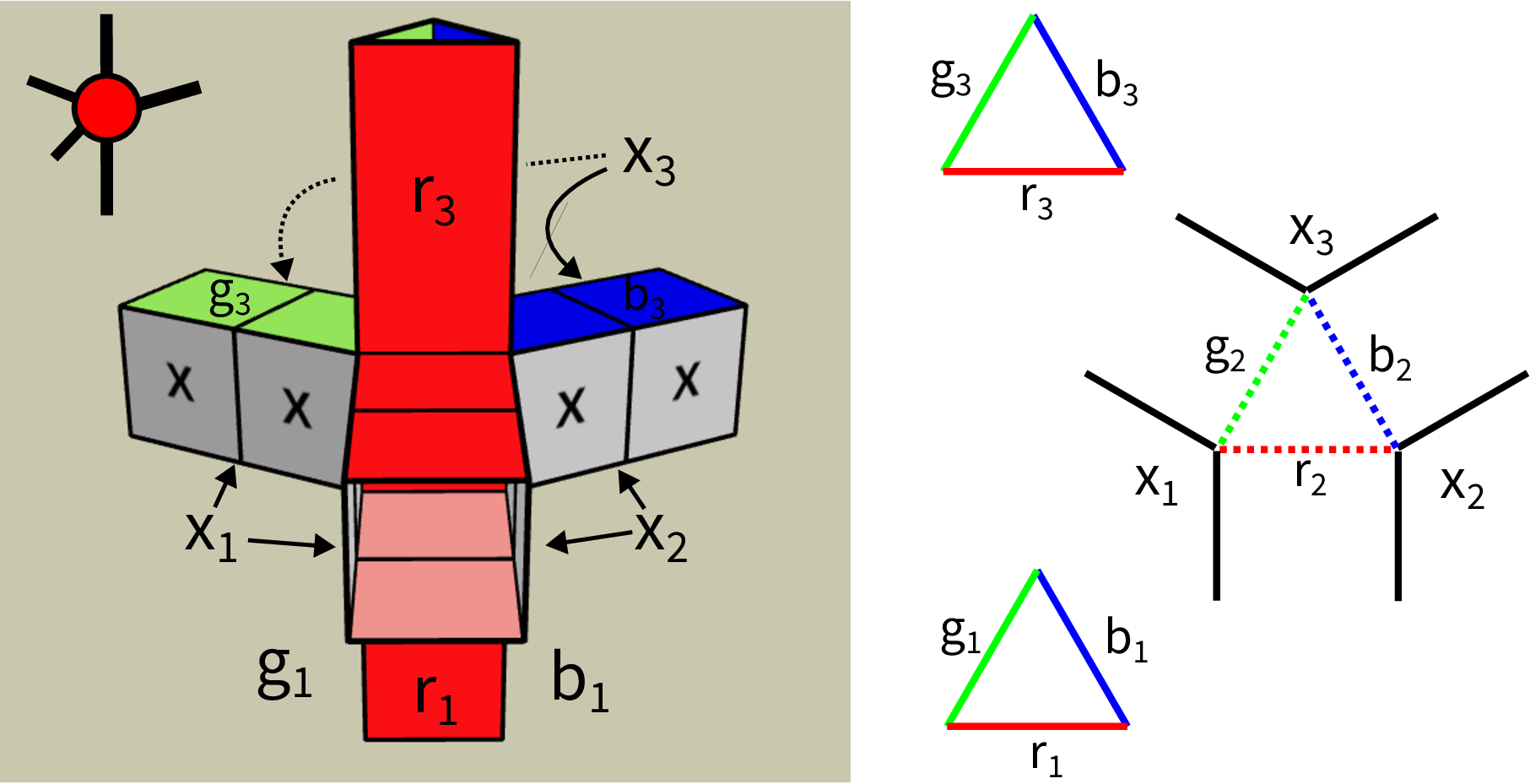}
  \caption{Left: A degree-5 junction in the spacetime layout of the 2D color code, with domain walls labelled. The lateral Pauli boundaries of the space pipes are all of type X, so this junction corresponds to an X spider in the ZX diagram as shown in the top left. Right: the cross sections at three heights. Solid lines represent opaque domain walls, while dashed lines represent semi-transparent domain walls.}
  \label{fig:zx-junction-with-color-boundary}
\end{figure*}

In this appendix we derive the junction--spider correspondence stated in \sref{sec:correspondence-zx}, by analyzing the propagation of logical Pauli operators through a junction. We take a degree-5 junction as an example, with all the space pipes having Pauli X boundaries laterally, as shown in \fref{fig:zx-junction-with-color-boundary}; the analysis generalizes directly to junctions of lower degree.

We first set up notation, referring to \fref{fig:zx-junction-with-color-boundary}. We slice the junction at three heights, below (1), at (2), and above (3) the space pipes. A colour label carries the height as a subscript. At heights 1 and 3 we label the colour boundaries of time pipes as $r_1,g_1,b_1$ and $r_3,g_3,b_3$. At height 2 we label the interface between the prism and space pipes as $r_2$, $g_2$ and $b_2$, where the domain wall $r_2$ condenses $\rx$, transmits the remaining bosons of the $\mathsf{x}$ row, $\gx$ and $\bx$, as the anyon $e$, transmits the remaining bosons of the $\mathsf{r}$ column, $\ry$ and $\rz$, as $m$, and confines the other four bosons; likewise for $g_2$ and $b_2$~\cite{Kesselring24Condensation}. Adjacent boundaries of the same type share one label: the bottom and top colour faces of the red space pipe adjoin $r_1$ and $r_3$ and inherit those labels, and the two lateral Pauli boundaries of neighbouring space pipes meeting at a vertical prism edge carry a common label $x_1$, $x_2$, or $x_3$. An endpoint of an $a$-string absorbed by a domain wall $W$ is denoted as $a|_W$, and a product of endpoint symbols denotes a single connected string operator with those endpoints. The label of a string changes where it crosses a semi-transparent wall, and we record only the endpoints. For two walls $W$ and $W'$, we denote their corner as $W\cap W'$.

A logical X operator is initialized as a star-shaped string operator in the bottom time pipe, and is then deformed in three steps:
\begin{equation}
\begin{aligned}
 &\rx|_{r_1}\times\bx|_{b_1}\times\gx|_{g_1} \\
={}&\rx|_{r_2}\times\bx|_{b_2}\times\gx|_{g_2} \\
={}&\rx|_{r_2}\times\rx|_{b_2\cap g_2} \\
={}&e|_{x_1}\times e|_{x_2}.
\end{aligned}
\label{eq:logical-x-propagation}
\end{equation}
First, the three endpoints slide upward from the colour boundaries onto the semi-transparent walls, which condense the same $\mathsf{x}$ bosons. Second, the $\bx$ and $\gx$ endpoints slide along their walls to the corner $b_2\cap g_2$, where the two legs fuse as $\bx\times\gx=\rx$; the operator is now a single $\rx$-string from $r_2$ to the corner. Third, each endpoint moves to a corner of the wall $g_2$ and crosses it, converting to $e$ inside the green space pipe and terminating on the adjacent lateral boundary; pushing the rest of the string through $g_2$ as well leaves an $e$-string connecting the two lateral boundaries $x_1$ and $x_2$ of the green pipe, which is the logical X operator of that pipe. A logical X entering from the bottom time pipe thus leaves the junction on exactly one other pipe. Similar deformations deliver it to any other single pipe, and products of such deformations produce every configuration in which an even number of the five pipes carries a logical X operator.

A logical Z operator is also initialized as a star in the bottom time pipe, but it cannot terminate anywhere at the junction height: the semi-transparent walls condense only $\mathsf{x}$ bosons, and the lateral faces of the pipes condense $e$. Instead, note that a $\rz$-string can run vertically through the prism bulk from $r_1$ to $r_3$; we write this through-going string as $\bigl[\rz|_{r_1}\times\rz|_{r_3}\bigr]$, and inserting its square multiplies the operator by the identity. Doing so for each colour and regrouping,
\begin{equation}
\begin{aligned}
 &\rz|_{r_1}\times\bz|_{b_1}\times\gz|_{g_1} \\
={}&\rz|_{r_1}\times\bigl[\rz|_{r_1}\times\rz|_{r_3}\bigr]^2 \\
 &\times\bz|_{b_1}\times\bigl[\bz|_{b_1}\times\bz|_{b_3}\bigr]^2 \\
 &\times\gz|_{g_1}\times\bigl[\gz|_{g_1}\times\gz|_{g_3}\bigr]^2 \\
={}&\bigl[\rz|_{r_3}\times\bz|_{b_3}\times\gz|_{g_3}\bigr] \\
 &\times\bigl[\rz|_{r_1}\times\rz|_{r_3}\bigr]
 \times\bigl[\bz|_{b_1}\times\bz|_{b_3}\bigr]
 \times\bigl[\gz|_{g_1}\times\gz|_{g_3}\bigr] \\
={}&\bigl[\rz|_{r_3}\times\bz|_{b_3}\times\gz|_{g_3}\bigr] \\
 &\times\bigl[m|_{r_1}\times m|_{r_3}\bigr]
 \times\bigl[m|_{b_1}\times m|_{b_3}\bigr]
 \times\bigl[m|_{g_1}\times m|_{g_3}\bigr].
\end{aligned}
\label{eq:logical-z-propagation}
\end{equation}
In the second step, each leg of the star is concatenated with one inserted string at their shared endpoint on the bottom boundary, extending the leg to the top boundary: the first bracket is again a star, now in the top time pipe, and one through-going string per colour is left over. In the last step, each leftover string is pushed sideways through the semi-transparent wall of its colour into the corresponding space pipe, converting the $\mathsf{z}$ boson to $m$; its endpoints land on the bottom and top colour boundaries of the pipe, which share the labels of the boundaries they adjoin. Each bracket in the final line is therefore an $m$-string connecting the bottom and top faces of one space pipe, which is the logical Z operator of that pipe. A logical Z entering from the bottom time pipe thus leaves the junction on all four other pipes simultaneously.

This propagation of logical operators is exactly the Pauli web through a phase-0 X spider: the $X$ operators on its wires obey the parity rule, and the $Z$ operators obey the all-or-nothing rule. Therefore, such a junction directly corresponds to a degree-5 X spider in the ZX diagram. The same analysis can be applied to a junction with Pauli Z boundaries laterally, which corresponds to a Z spider in the ZX diagram.

Note that in \eref{eq:logical-z-propagation}, an initial logical Y operator can also transform into 3 $m$-strings in the three space pipes, which may seem to violate the Pauli web of a ZX diagram. The root cause is that a color boundary in the spacetime layout is completely symmetric for the three Pauli labels~\cite{kesselringBoundariesTwistDefects2018}, but the ZX diagram is not symmetric. In this case, we have to choose a specific identification of the $m$-strings in the three space pipes as a specific Pauli label, e.g., Z operator, to make the correspondence to a ZX diagram.

Equations~\eqref{eq:logical-x-propagation} and \eqref{eq:logical-z-propagation} also hold if we use Pauli boundaries for the top and bottom faces of the space pipe, instead of the colour boundaries. This breaks the symmetry of the three Pauli labels more explicitly, and makes the identification of anyon strings in the space pipes as a specific Pauli label more natural. Macroscopically, both choices of the top and bottom faces of the space pipe are equivalent, and do not affect the correspondence to ZX diagram, but can affect the microscopic circuit on the physical qubits.

\bibliographystyle{quantum}
\bibliography{references}
\end{document}